\documentclass{aa}
\usepackage{graphicx}
\usepackage{txfonts}
\begin{document}
   \title{On the incidence rate of first overtone Blazhko stars 
          in the Large Magellanic Cloud}

   \author{A. Nagy\inst{} and G. Kov\'acs\inst{}}


   \institute{
   Konkoly Observatory, P.O. Box 67, H-1525,
   Budapest, Hungary \\ \email {nagya, kovacs@konkoly.hu}
   \\
   }

   \date{Received / Accepted }

%
%
   \titlerunning {Incidence rate of Blazhko stars}
   \abstract
{} 
{By using the full span of multicolor data on a representative 
sample of first overtone RR~Lyrae stars in the Large Magellanic 
Cloud (LMC) we revisit the problem of the incidence rate of 
the amplitude/phase-modulated (Blazhko) stars.}
{Multicolor data, obtained by the MAssive Compact Halo Objects 
(MACHO) project, are utilized through a periodogram averaging 
method.} 
{The method of analysis enabled us to increase the number of 
detected multiperiodic variables by $18$\% relative to the 
number obtained by the analysis of the best single color data. 
We also test the maximum modulation period detectable in the 
present dataset. We find that variables showing amplitude/phase 
modulations with periods close to the total time span can still 
be clearly separated from the class of stars showing period 
changes. This larger limit on the modulation period, the more 
efficient data analysis and the longer time span lead to a 
substantial increase in the incidence rate of the Blazhko stars 
in comparison with earlier results. We find altogether 99 first 
overtone Blazhko stars in the full sample of 1332 stars, implying 
an incidence rate of $7.5$\%. Although this rate is nearly twice 
of the one derived earlier, it is still significantly lower than 
that of the fundamental mode stars in the LMC. The by-products 
of the analysis (e.g., star-by-star comments, distribution 
functions of various quantities) are also presented.} 
{}


   \maketitle
%

%
%
\section{Introduction}
Current analyses of the databases of the microlensing projects 
MACHO and OGLE (Optical Gravitational Lensing Experiment) have 
led to a substantial increase in our knowledge on the frequency 
of the amplitude/phase-modulated RR~Lyrae (Blazhko [BL]
\footnote{See Szeidl \& Koll\'ath (2000) for a historically 
more precise possible nomenclature.}) stars. Moskalik \& Poretti 
(2003) analyzed the OGLE-I data on 215 RR~Lyrae stars in the 
Galactic Bulge. They found incidence rates for the fundamental 
(RR0) and first overtone (RR1) Blazhko stars of $23$\% and $5$\%, 
respectively. A similar analysis by Mizerski (2003) on a much 
larger dataset from the OGLE-II database yielded $20$\% and $7$\%, 
respectively. These rates, at least for the RR0 stars, are very 
similar to the ones suggested some time ago by Szeidl (1988) 
from the various past analyses of rather limited data on Galactic 
field and globular cluster stars. One gets smaller rates by 
analyzing the RR~Lyrae stars in the Magellanic Clouds. Based 
on the OGLE-II observations of a sample of 514 stars in the 
Small Magellanic Cloud (SMC), in a preliminary study, 
Soszy\'nski et al. (2002) obtained the same rate of $10$\% both 
for the RR0 and RR1 stars. In a similar study on a very large 
sample containing 7110 RR~Lyrae stars in the LMC, Soszy\'nski et al. 
(2003) derived $15$\% and $6$\% for the RR0 and RR1 stars, 
respectively. In earlier studies on the same galaxy, based on 
the observations of the MACHO project, Alcock et al. (2000, 2003, 
hereafter A00 and A03, respectively) got rates of $12$\% and $4$\% 
for the above two classes of variables. One may attempt to 
relate these incidence rates to the metallicities of the various 
populations (e.g., Moskalik \& Poretti 2003), but the relation 
(if it exists) is certainly not a simple one (Kov\'acs 2005; 
Smolec 2005). 

Except for the SMC, all investigations indicate much lower 
incidence rates for RR1$-$BL stars than for RR0$-$BL ones. 
It is believed that the cause of this difference is internal, 
i.e., due to real difference in physics and can not be 
fully accounted for by the possibly smaller modulation amplitudes 
of the RR1 stars. Since this observation may have important 
consequences on any future modeling of the BL phenomenon, 
we decided to re-analyze the MACHO database and utilize all 
available observations (i.e., full time span two color data). 
In Sect.~2 we summarize the basic parameters of the datasets 
and some details of the analysis. Section~3 describes our method 
for frequency spectrum averaging. The important question of the 
longest detectable BL period from the present dataset and the 
concomitant problem of variable classification are dealt with 
in Sect.~4. Analysis of the RR1 stars with their resulting 
classifications will be presented in Sect.~5. Finally, in Sect.~6 
we summarize our main results with a brief discussion of the 
current state of the field.      

%
%
\section{Data, method of analysis}
For comparative purpose, in the first part of our analysis, we 
use the same dataset as the one employed by A00. Our final results 
are based on the full dataset spanning $\sim 7.5$~years. Basic 
properties of these two sets are listed in Table~1. Both sets 
contain the same $1354$ stars and cover the fields \# 2, 3, 5, 6, 
9, 10, 11, 12, 13, 14, 15, 18, 19, 47, 80, 81 and 82, sampling 
basically the LMC bar region. Because the selection of the stars 
was made earlier on simple preliminary criteria such as period 
and color, some variables, other than first overtone RR~Lyrae 
stars, were also included. Fortunately, the number of these other 
variables is only 22, that is small relative to the full sample.  

%
%
\begin{table}
\caption{Properties of the LMC RR1 datasets analyzed in this paper}
\begin{flushleft}
\begin{tabular}{cccc}
\hline\hline
{\rm Set} & $\langle T_{\rm tot} \rangle$ & $\langle N_{\rm d} \rangle$ 
& Colors \\
\hline
\#1   & $6.5$   & $700$   & `r', `b' \\
\#2   & $7.5$   & $900$   & `r', `b' \\
\hline
\end{tabular}
\end{flushleft}
{\footnotesize
\underline {Notes:} 
$\langle T_{\rm tot} \rangle = $~average total time span [yr];  
$\langle N_{\rm d} \rangle = $~average number of datapoints per variable; 
Colors: MACHO instrumental magnitudes, see Alcock et al. (1999)
}
\end{table}

We employed a standard Discrete Fourier Transform method 
by following the implementation of Kurtz (1985). All analyses 
were performed in the $[0,6]$~d$^{-1}$ band with $150000$ frequency 
steps, ensuring an ample sampling of the spectrum line profiles 
even for the longest time series. The search for the secondary 
frequencies was conducted through successive prewhitenings in the 
time domain. The first step of it consisted of the subtraction 
of the main pulsation component together with its harmonics up 
to order three. In all cases the frequency of the given component 
was made more accurate by direct least squares minimization. 
When the two colors were used simultaneously, we allowed different 
amplitudes and phases for the two colors and minimized the harmonic 
mean of the standard deviations of the respective residuals. 
Although somewhat arbitrarily, we considered the harmonic mean 
as a useful function for taking into account differences in the 
data quality.

For the characterization of the signal-to-noise ratio (SNR) of 
the frequency spectra we use the following expression 
%
%
\begin{eqnarray}
{\rm SNR} = {A_{\rm p}-\langle A_{\nu}\rangle\over \sigma_{A_{\nu}}} 
\hskip 2mm ,
\end{eqnarray}
where $A_{\rm p}$ is the amplitude at the highest peak in the 
spectrum, $\langle A_{\nu}\rangle$ is the average of the spectrum 
and $\sigma_{A_{\nu}}$ is its standard deviation, computed by an 
iterative $4\sigma$ clipping.

%
%
\section{Spectrum Averaging Method (SAM)}
Unlike OGLE, where measurements are taken mostly in the $I_c$ 
band (e.g., Udalski, Szyma\'nski \& Kubiak, M. 1997), the 
MACHO database has roughly equal number of measurements in 
two different colors for the overwhelming majority of the 
objects (e.g., Alcock et al. 1999). This property of the 
MACHO database enables us to devise a method that uses both 
colors for increasing the signal detection probability.

It is clear that simple averaging of the time series of the 
different color bands cannot work, because: (a) the two time 
series are systematically different due to the difference in 
colors; (b) sampling may be different for hardware, weather or
other reasons. Therefore, we are resorted to some spectrum 
averaging method that uses the frequency spectra of the two 
time series rather than their directly measured values. 
Obviously, SAM is also affected by the time series properties 
mentioned above. However, this effect is a weaker, because: 
(i) we are interested in the positions of the peaks in the 
spectra and these are the same whatever colors are used; 
(ii) spectral windows are similar, unless the samplings of 
the two time series are drastically different. As an example 
for the fulfilment of this latter condition for the MACHO 
data, in Fig.~1 we show the spectral windows in the two colors 
in one representative case. 
%
%
%
   \begin{figure}[h]
   \centering
   \includegraphics[width=70mm]{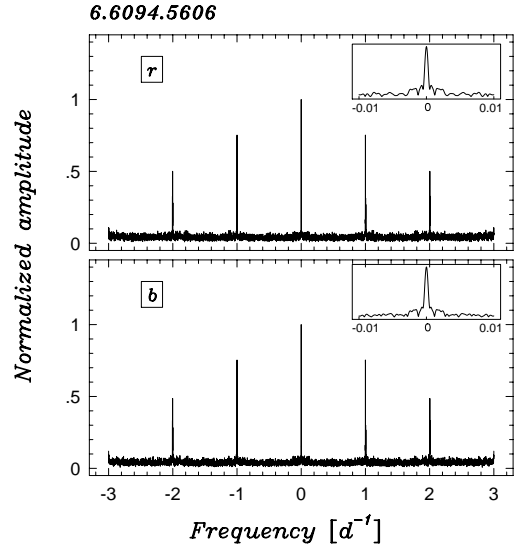}
      \caption{Example of the similarity of the spectral windows 
               of the MACHO instrumental `r' and `b' time series. 
	       Close-ups of the main peaks are displayed in the 
	       insets. The MACHO identification number of the star 
	       analyzed is shown in the upper left corner of the 
	       figure.} 
         \label{fig1}
   \end{figure}

In order to optimize noise suppressing, we compute the summed spectrum 
by weighting the individual spectra by the inverse of their variances 
%
%
\begin{eqnarray}
S_i={\sigma_{\rm r}^2\sigma_{\rm b}^2 \over \sigma_{\rm r}^2+\sigma_{\rm b}^2}
\Bigl({1\over \sigma_{\rm r}^2}R_i+{1\over \sigma_{\rm b}^2}B_i\Bigr)
\hskip 2mm ,
\end{eqnarray}
where $R_i$ and $B_i$ are the amplitude spectra of the `r' and `b' 
data, $\sigma_{\rm r}$ and $\sigma_{\rm b}$ are the standard deviations 
of the spectra. 

Before we discuss the signal detection capability of SAM, it is 
necessary to give significance levels for periodic signal detection 
when using the present dataset. Perhaps the simplest way of doing 
this is to perform a large number of numerical simulations and 
derive an empirical distribution function of SNR. For this goal 
we generated pure Gaussian noise on the observed time base of 10 
randomly selected stars. For each object we generated 1000 different 
realizations and computed the frequency spectra in our standard 
frequency band of $[0,6]$~d$^{-1}$. The empirical distribution 
function was derived from the 10000 SNR values computed on these 
spectra. Figure~2 shows the resulting functions for the single 
color and SAM spectra. For further reference, the 1\% significance 
levels are at $6.9$ and $6.5$ for the single color and SAM spectra, 
respectively. All signals that have lower SNR values than these, 
are considered to be {\it not detected} in the given time series. 
The lower cutoff for the SAM spectra is the result of the averaging 
of independent spectra which also results in a greater number of 
independent points in the SAM spectra and concomitantly, in a 
slightly different shape of the distribution functions (see also 
Kov\'acs, Zucker \& Mazeh 2002).
%
%
%
   \begin{figure}[h]
   \centering
   \includegraphics[width=80mm]{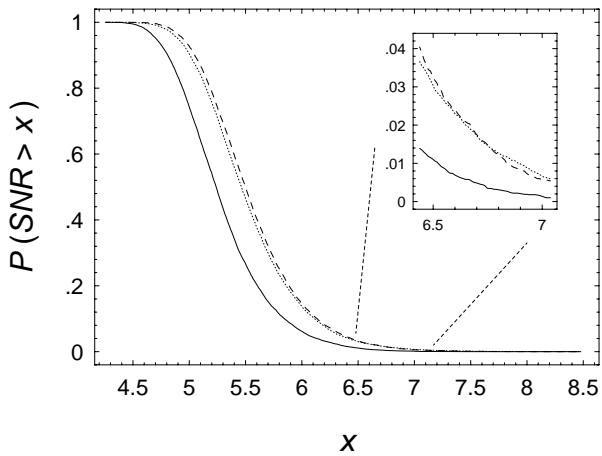}
      \caption{
Probability distribution functions of SNRs of the amplitude spectra  
of pure Gaussian noise generated on the time base of the `r' data 
(dotted line), `b' data (dashed line) and their combination by SAM 
(solid line). 
      } 
         \label{fig2}
   \end{figure}

In turning to the signal detection efficiency of SAM, we note the 
following. At very low SNR we expect no improvement by applying 
SAM, because the optimally achievable noise suppression is 
insufficient for the signal to emerge from the noise. In the 
other extreme, when the the signal is strong, the improvement is 
expected to be minimal and secure detection is possible without 
employing SAM. In order to verify this scenario, we conducted 
tests with artificial data. The result of one of these tests is 
shown in Fig.~3. The test data were generated on the 7.5~year 
time base given by the variable 10.3434.936. The signal consisted 
of a simple sinusoidal with a period of $0.278$~d and amplitude 
$A$, where $A$ was chosen to be the same in both colors and 
changed in $100$ steps between $0.01$ and $0.3$ to scan different 
SNR values. At each value of the amplitude we added Gaussian noise 
of $\sigma=0.11$ to the signal. This procedure was repeated for 
$100$ different realizations. At each color and amplitude we computed 
the average and standard deviation of the SNR values obtained from 
these $100$ realizations. In order to characterize the gain by 
employing SAM, we also computed the ratio of the SNR values 
($R_{\rm SNR}$) as derived from the single color data and by the 
application of SAM.    
%
%
%
   \begin{figure}[h]
   \centering
   \includegraphics[width=80mm]{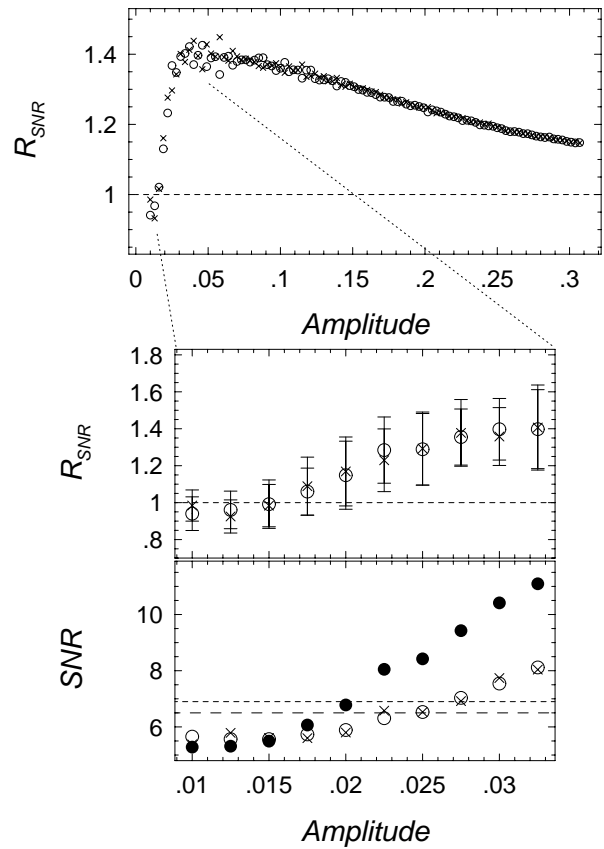}
      \caption{
{\it Uppermost panel:} variation of the ensemble average of 
$R_{\rm SNR}={\rm SNR}_{\rm SAM}/{\rm SNR}_{\rm color}$ as a 
function of the test amplitude. Open circles refer to the `r' 
data, crosses to the `b' data. {\it Middle panel:} close-up 
of the low/mid SNR region of the above diagram. Error bars 
show the $1\sigma$ ranges of the values obtained by the various 
noise realizations. {\it Lowermost panel:} ensemble average 
of SNR. The dots are for the SAM values. Short- and long-dashed 
lines show the $1$\% noise probability levels at SNR~$=6.9$ 
and SNR~$=6.5$, corresponding to the single-color and SAM 
spectra, respectively (see also text and Fig.~2 for further 
details).      
      } 
         \label{fig3}
   \end{figure}

We see from Fig.~3 that there is a region where the signal 
has too low amplitude to be detected in the single color data, 
but not low enough to remain hidden in the SAM spectra. This 
is the most interesting amplitude/noise regime, where the 
application of SAM results in new discoveries. At higher 
amplitudes SAM leads `only' to an increase in SNR. It is seen 
that the maximum increase in SNR is at around $\sqrt 2$, as 
expected from elementary considerations. 

In order to show non-test examples for the signal detection 
efficiency of SAM, in Figs.~4 and 5 we exhibit the method at 
work for the observed data of two variables. In both cases the 
single color data are insufficient for detection, but the 
combined spectra show clearly the presence of the signal.  
%
%
   \begin{figure}[h]
   \centering
   \includegraphics[width=70mm]{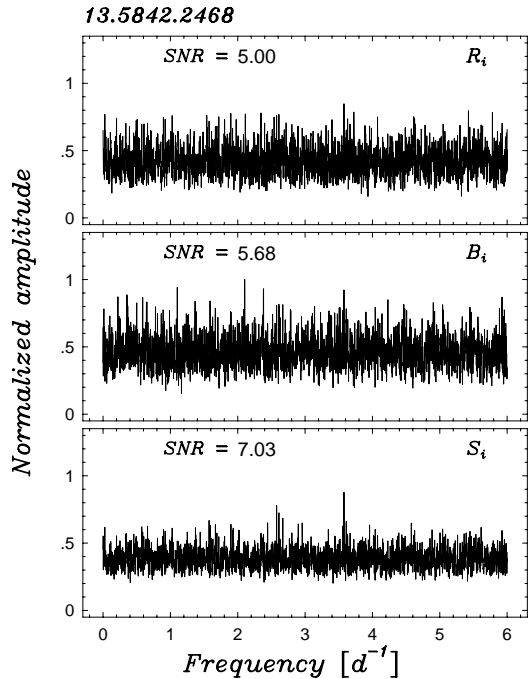}
      \caption{
Example of the signal detection efficiency of SAM at low SNR level, 
when the single color data do not show the presence of the signal. 
Upper and middle panels show the frequency spectra for the `r' and 
`b' data, whereas the lower one exhibits the SAM result. The MACHO 
star identification number is shown at the top, whereas the SNR 
values are given in the corresponding panels. All spectra are 
displayed in the same (arbitrary) scale.     
      } 
         \label{fig4}
   \end{figure}
%
%

%
%
   \begin{figure}[h]
   \centering
   \includegraphics[width=70mm]{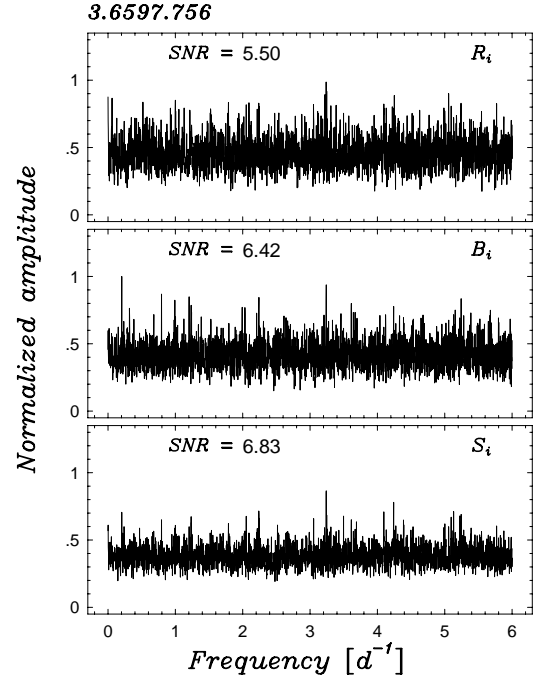}
      \caption{
As in Fig.~4, but for the case when the signal is at the verge of 
detection in the single color data and SAM increases the reliability 
of the detection.  
      } 
         \label{fig5}
   \end{figure}

By analyzing all the 1354 variables of set \#1 we can count the 
number of cases when significant components are found after 
the first prewhitening. The result of this exercise is shown in 
Table~2. We see that there is a strong support of the obvious 
assumption that application of two color data significantly 
increases the detection probability of faint signals. The gain 
is 18\% relative to the best single color detection rate.  

%
%
\begin{table}
\caption{Number of single- and multi-periodic variables in set 
\#1 of Table~1.}
\begin{flushleft}
\begin{tabular}{lrrr}
\hline\hline
{\rm Color} & r & b & SAM\\
\hline
$N_1          $  & $978 $   & $895 $   & $813 $   \\
$N_2          $  & $376 $   & $459 $   & $541 $   \\
$N_2/(N_1+N_2)$  & $27.8$\% & $33.9$\% & $40.0$\% \\
\hline
\end{tabular}
\end{flushleft}
{\footnotesize
\underline {Notes:} $N_1=$~number of single-periodic variables; 
$N_2=$~number of multi-periodic variables
}
\end{table}

%
%
\section{The phenomenological definition of the BL stars 
{\it -- Separation of the BL and PC variables}
}
The current loose definition of the BL stars relies purely 
on the type of the frequency spectra of their light or radial 
velocity variations. Usually those RR~Lyrae stars are called 
BL variables that exhibit closely spaced peaks in their 
frequency spectra. Although most of the variables classified 
as type BL contain only one or two symmetrically spaced 
peaks at the main pulsation component, complications arise 
when, after additional prewhitenings, significant components 
remain, or, when the prewhitening is ambiguous, due to the 
closeness of the secondary components. There are two basic 
cases of complication: 
(i) the remnant components are not well-separated and 
few successive prewhitenings do not lead to a complete 
elimination of the secondary components; 
(ii) we have several well-defined additional peaks, whose 
pattern is different from the ones usually associated with 
the BL phenomenon (single-sided for BL1 and symmetrically 
spaced for BL2 patterns).    

Case (i) can be considered as a consequence of some 
non-stationarity in the data, leading to a nearly continuous 
spectral representation, whereas case (ii) can be related 
perhaps to some modulation of more complicated type than 
that of the BL phenomenon. In order to make the phenomenological 
classification of variable types as clear as possible, 
supported by the test to be presented below, we adopt the 
following definitions:

\begin{itemize}
\item[]
\begin{itemize}
\item[\underline{PC}] 
(period changing stars): 
Close components (i.e., the ones that are close to the main 
pulsation frequency) cannot be eliminated within three 
prewhitenening cycles, or, if they can, their separation 
from the main pulsation component is less than $\sim 1/T$, 
where $T$ is the total time span. Furthermore, except for 
possible harmonics, there are no other significant components 
in the spectra.  
\end{itemize}

\item[]
\begin{itemize}
\item[\underline{BL}] 
(Blazhko stars):
There is either only one or two symmetrically spaced close 
components on both sides of the main pulsation frequency. 
After the second or third prewhitenings, except for possible 
harmonics, no significant pattern remains or the pattern is 
of type PC.
\end{itemize}

\item[]
\begin{itemize}
\item[\underline{MC}] 
(stars with closely spaced multiple frequency components): 
All close components are well-separated (i.e., frequency distances 
are greater than $\sim 1/T$) and, except if there are only two 
secondary components on one side of the main pulsation frequency, 
more than three prewhitenings are necessary to eliminate all of them.
\end{itemize}
\end{itemize}    

We note that possible appearance of instrumental effects (i.e., 
peaks at integer d$^{-1}$ frequencies) is disregarded in the above 
classification. Class MC is equivalent to class $\nu$M used in A00 
and A03. Here the new notation is merely aimed at simplification 
on the occasion of the above definition. Furthermore, in the 
previous works, variables with ``dominantly'' BL2 structures were 
included in class BL2. Now the above scheme puts them among the 
MC stars.

Based on the following test, we note that below frequency 
separations less than $\sim 1.5/T$, distinction among the 
above types becomes more ambiguous, because in the simple 
prewhitening technique followed in this paper, the result 
will depend on the phase of the modulation.  

In order to substantiate the above definitions, here we examine 
in more detail how we can distinguish between the PC and BL 
phenomena based solely on the properties of the prewhitened spectra.  

We use the time distribution of arbitrarily selected stars to 
generate artificial time series with modulated signal parameters 
given in Table~3. We see that the chosen types of modulation  
cover nearly all basic cases in the lowest order approximation 
(i.e., single main pulsation component with frequency $\omega_0$, 
linear period change, etc.). No noise is added, because we are 
interested in differences caused by the various non-stationary 
components in the residual spectra, after prewhitened by the 
pulsation component. Nevertheless, noise plays an important 
role at low modulation levels, but a more detailed study of the 
complicated problem of non-stationary signal classification in 
the presence of noise is out of the scope of the present work.
   
%
%
\begin{table}
\caption{Definition of the signals used in the BL/PC variability test.}
\begin{flushleft}
\begin{tabular}{ll}
\hline\hline
Type & Time dependence \\
\hline
AM &  $A(t) = A\sin(\Omega t + \Phi)$; 
\hfill $\omega(t)$, $\varphi(t)=$~const. \\
PM &  $\varphi(t) = A_{\varphi}\sin(\Omega t + \Psi)$; 
\hfill $\omega(t)$, $A(t)=$~const.\\
FM &  $\omega(t)  = \omega_0+A_{\omega}\sin(\Omega t + \Gamma)$; 
\hfill $A(t)$, $\varphi(t)=$~const.\\
PC         &  $\omega(t)  = \omega_0/(1+\beta t/P_0)$; 
\hfill $A(t)$, $\varphi(t)=$~const.\\
\hline
\end{tabular}
\end{flushleft}
{\footnotesize
\underline {Notes:} Signal form: $x(t)=(1+A(t))\sin(\omega(t)t+\varphi(t))$; 
$\omega_0=2\pi/P_0$; $P_0=0.377$; $\Omega=2\pi/P_{\rm BL}$; 
$P_{\rm BL} =$~BL period; 
AM=amplitude modulation; 
PM=phase modulation;
FM=frequency modulation;
PC=secular frequency change; 
$\Phi$, $\Psi$, $\Gamma$ are arbitrary constant phases.
}
\end{table}

The same pulsation period of $0.377$~d is used for 10 randomly 
selected stars. For each star, we scan the modulation frequency 
$\Omega$ and the rate of period change $\beta$ at fixed values 
of all the other parameters. Amplitudes are adjusted to get 
modulation levels that are 20\% of the peak of the main pulsation 
component in the frequency spectra. Parameter $\beta$ is changed 
in the range of $(1$--$12)\times 10^{-8}$, yielding modulation 
levels between $20$\% and $50$\%. Each scan is repeated with 
10 randomly selected phase values. By using the same code for 
the analysis of these artificial signals as the one employed 
for the observed data, we get the result shown in Fig.~6. In 
order to characterize the efficiency of the prewhitening, here 
we use the ratio $A_3/A_1$, where $A_1$ and $A_3$ denote the 
peak amplitudes after the first and third prewhitenings, 
respectively. Although in practice we use the SNR of the 
frequency spectra to select significant components, here the 
direct comparison of the amplitudes is more meaningful, because 
for noiseless signals SNR can be high even if the peak amplitude 
itself is small.      
%
%
%
   \begin{figure}[h]
   \centering
   \includegraphics[width=80mm]{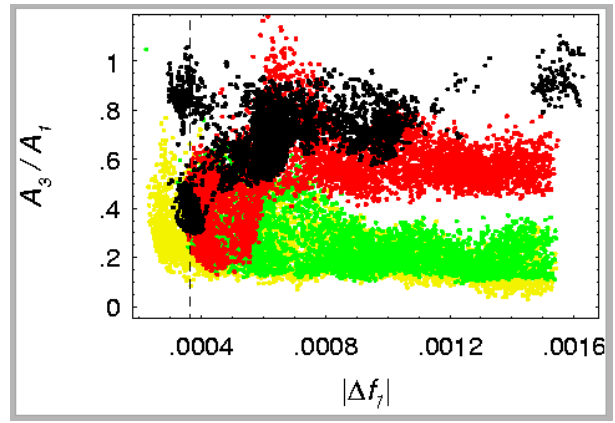}
      \caption{Testing various modulated signals as given in Table~3. 
      The ratio of the peak amplitudes after the third and first 
      prewhitenings are plotted as a function of the measured 
      frequency distance (in [d$^{-1}$]) from the main pulsation 
      component. Dashed vertical line shows the position of the $1/T$ 
      frequency, where $T$ denotes the total time span. Dots of 
      different shades/colors are for the various modulation types: 
      yellow/light gray, green/gray, red/dark gray and black are 
      for AM, PM, FM and PC, respectively. The separation of simple 
      amplitude and phase modulations from the more complicated 
      ones (FM and PC) are clearly visible at $|\Delta f_1|> 2/T$.
      } 
         \label{fig6}
   \end{figure}

It is clear from the figure that PC signals remain difficult 
to prewhiten, except perhaps near and below $1/T$, when any 
types of signal can be prewhitened due to the proximity to the 
main pulsation component. Although periodic frequency modulation 
behaves similarly to secular period change above $1.5/T$, it 
can be prewhitened (and therefore confused) with simple amplitude 
and phase modulations below this limit. When the secular change 
or periodic modulation of the pulsation frequency is strong, 
they are more easily separated from simple modulated signals 
(of types AM and PM, see Table~3). This happens at measured 
modulation frequencies greater than $\sim 2/T$ (see the gap 
starting at $\sim 0.0008$~d$^{-1}$). Additional separation 
between AM and PM signals is observable at modulation frequencies 
greater than $\sim 5/T$ , where, as expected, the AM type leaves 
the weakest trace after the third prewhitening. 

We conclude that the standard method of prewhitening we use for 
the analysis of the observed data is capable of making distinction 
between linear period change and signal modulation, but confusion 
occurs with periodic frequency modulations under observed 
modulation frequencies of $1.5/T$. Modulation types AM and FM 
are also difficult to distinguish, except for short modulation 
periods, when type AM can be prewhitened more easily. These 
results support our phenomenological classification of variables 
given at the beginning of this section even in the case of 
long modulation periods.

%
%
\section{New detections by using SAM}
Here we re-address the question of the incidence rate of first 
overtone BL stars by applying the method of analysis and scheme 
of classification described in the previous sections.
%
%
\begin{table*}[h]
\caption{Final classification of the 1354 pre-selected variables}
\begin{flushleft}
\begin{tabular}{llrr}
\hline\hline
Classification & Short description & Number & Inc. rate in RR1 \\
\hline
RR1--S    & Singly-periodic overtone RR Lyrae & $712$ & $53.5\,$\% \\
RR1--BL1  & RR1 with one close component      & $46 $ & $3.5\, $\% \\
RR1--BL2  & RR1 with symmetric frequencies    & $53 $ & $4.0\, $\% \\
RR1--MC   & RR1 with more close components    & $13 $ & $1.0\, $\% \\
RR1--PC   & RR1 with period change            & $187$ & $14.0\,$\% \\
RR1--D    & RR1 with frequencies at integer 
            $d^{-1}$                          & $137$ & $10.3\,$\% \\
RR1--MI   & RR1 with some miscellany          & $13 $ & $1.0\, $\% \\
\hline
RR01      & FU/FO double-mode RR Lyrae        & $165$ & $12.4\,$\% \\
RR01--BL1 &                                   & $1  $ & $0.1\, $\% \\
RR01--PC  &                                   & $5  $ & $0.4\, $\% \\
\hline
RR0--S    & Fundamental mode RR Lyrae         & $1  $ & $-$        \\
RR0--BL1  &                                   & $1  $ & $-$        \\
RR0--BL2  &                                   & $1  $ & $-$        \\
\hline
MDM       & Mysterious double-mode            & $3  $ & $-$        \\
BI        & Eclipsing binary                  & $16 $ & $-$        \\
\hline
\end{tabular}
\end{flushleft}
\end{table*}
%
%

%
%
\begin{table*}[h]
\caption{Notebook of the analysis}
\begin{flushleft}
\begin{tabular}{lcllll}
\hline\hline
MACHO ID&Period&Type $r$&Type $r,\,b$&Type $r,\,b$&Comments\\
&&(A00)&(SAM)&(SAM)&\\
&[d]&\multicolumn{2}{c}{6.5 years}&7.5 years&\\
\hline
10.3190.501 &    0.3527946&    RR1-S&      RR1-S&      RR1-S&\\
10.3191.363  &   0.4060709 &   RR01  &     RR01  &     RR01&\\
10.3193.457   &  0.2977060  &  RR1-S  &    RR1-S  &    RR1-S&      
weak peak in r at 0.4477 \\
10.3193.514&     0.2762716   & RR1-S   &   RR1-S   &   RR1-S&\\
10.3310.723 &    0.3675948&    RR1-S    &  RR1-S    &  RR1-S&\\
10.3311.612  &   0.3308571 &   RR1-S     & RR1-S     & RR1-S&\\
10.3314.787   &  0.3575062  &  RR1-S      &RR1-S      &RR1-S&\\
10.3314.873    & 0.3842321   & RR1-S&      RR1-S&      RR1-S&\\
10.3314.916&     0.3438987    &RR1-S &     RR1-S &     RR1-S&\\
10.3315.758 &    0.2755685&    RR1-S  &    RR1-S  &    RR1-S&\\
10.3432.666  &   0.3418437 &   RR1-S   &   RR1-PC  &   RR1-PC&\\
10.3434.825   &  0.2776860  &  RR1-S    &  RR1-S    &  RR1-S&\\
10.3434.936    & 0.2769524   & RR1-S     & RR1-S     & RR1-S&\\
10.3435.907&     0.3295068    &RR1-S      &RR1-S      &RR1-S&\\
10.3550.888 &    0.3317657&    RR1-S&      RR1-S&      RR1-S&\\
10.3552.745  &   0.2922940 &   RR1-BL1&    RR1-BL1&    RR1-BL1&\\
10.3556.986   &  0.2961494 &   RR1-PC  &   RR1-PC  &   RR1-PC  &   
df=0.00038\\
10.3557.1024   & 0.2947668  &  RR1-PC   &  RR1-PC   &  RR1-BL2  &  
df=0.00043\\
\hline
\end{tabular}
\end{flushleft}
{\footnotesize
\underline {Note:} This table is available in its entirety at 
CDS (http://cdsweb.u-strasbg.fr). Additional data (positions, 
magnitudes, etc.) can be found at the MACHO online database 
(http://wwwmacho.mcmaster.ca/Data/MachoData.html).

}
\end{table*}
%
%

%
%
%
\begin{table}
\caption{Properties of the MDM stars}
\begin{flushleft}
\begin{tabular}{llccc}
\hline\hline
{\rm MACHO ID} & $P_1$ & $P_2/P_1$ & $V$ & $V-R$ \\
\hline
12.10202.285   & 0.398 & 0.807 & 18.99 & 0.50 \\ 
12.10443.367   & 0.337 & 0.802 & 18.90 & 0.45 \\ 
9.4278.179     & 0.327 & 0.805 & 18.69 & 0.29 \\
\hline
\end{tabular}
\end{flushleft}
\end{table}    
%
%

%
%
\begin{table*}
\caption{Data on the first overtone Blazhko stars}
\tabcolsep 3.7pt
\begin{flushleft}
\begin{tabular}{lcrccc|lcrccc}
\hline\hline
MACHO ID.     &  $P_{1}$   &  $f_{\rm BL}$~~~~~& $A_+$  & $A_0$  & $A_-$  &MACHO ID.     &  $P_{1}$   &  $f_{\rm BL}$~~~~~& $A_+$  & $A_0$  & $A_-$  \\
\hline
10.3552.745   &  0.2922940 &     0.037408 & 0.0666 & 0.1280 &  $-$   &6.5729.958    &  0.2778179 &     0.074254 & 0.1045 & 0.1641 &  $-$   \\
10.3557.1024  &  0.2947668 &     0.000426 & 0.1000 & 0.2643 & 0.0941 &6.5730.4057   &  0.2763214 &     0.077598 & 0.0662 & 0.1249 &  $-$   \\
10.4035.1095  &  0.3188217 &     0.080780 & 0.0496 & 0.2239 &  $-$   &6.5850.1081   &  0.3335893 &  $-$0.118341 &  $-$   & 0.1617 & 0.0326 \\
10.4161.1053  &  0.2874579 &     0.094288 & 0.0769 & 0.1443 &  $-$   &6.5971.1233   &  0.2877179 &  $-$0.073476 & 0.0380 & 0.1868 & 0.0666 \\
11.9471.780   &  0.2859825 &  $-$0.113512 &  $-$   & 0.2076 & 0.0764 &6.6091.1198   &  0.2668322 &  $-$0.075389 &  $-$   & 0.1654 & 0.0307 \\
13.5713.590   &  0.2836353 &     0.100165 & 0.0636 & 0.0870 &  $-$   &6.6091.877    &  0.3206106 &  $-$0.038525 &  $-$   & 0.1046 & 0.0642 \\
13.5714.442   &  0.3168726 &     0.109252 & 0.0547 & 0.1008 &  $-$   &6.6094.5606   &  0.3393854 &  $-$0.001244 & 0.0593 & 0.2476 & 0.0637 \\
13.5842.2468  &  0.2731424 &  $-$0.084060 &  $-$   & 0.2060 & 0.0355 &6.6326.424    &  0.3304519 &  $-$0.072633 &  $-$   & 0.2353 & 0.0504 \\
13.5959.584   &  0.3466458 &     0.000993 & 0.0286 & 0.2629 &  $-$   &6.6810.616    &  0.2883850 &  $-$0.000403 & 0.0552 & 0.2113 & 0.0704 \\
13.6322.342   &  0.2621495 &     0.180083 & 0.0385 & 0.1166 &  $-$   &6.7054.713    &  0.4404073 &  $-$0.001674 & 0.0445 & 0.2110 & 0.0478 \\
13.6326.2765  &  0.3304520 &  $-$0.072702 &  $-$   & 0.2387 & 0.0334 &6.7056.836    &  0.4476160 &     0.001120 & 0.0317 & 0.1781 & 0.0252 \\
13.6810.2981  &  0.2883863 &     0.000419 & 0.0691 & 0.2050 & 0.0659 &80.6352.1495  &  0.3079889 &  $-$0.071798 &  $-$   & 0.0937 & 0.0682 \\
13.6810.2992  &  0.2903975 &  $-$0.041503 & 0.0909 & 0.2153 & 0.1318 &80.6470.2000  &  0.3706713 &     0.000564 & 0.0492 & 0.2016 & 0.0472 \\
14.8495.582   &  0.2971105 &     0.044128 & 0.0696 & 0.2236 & 0.0364 &80.6595.1599  &  0.3323705 &  $-$0.036844 & 0.0326 & 0.1977 & 0.0812 \\
14.9223.737   &  0.3098789 &     0.031590 & 0.0941 & 0.2270 & 0.1075 &80.6597.4435  &  0.3086445 &  $-$0.000950 & 0.0368 & 0.1503 & 0.0519 \\
14.9225.776   &  0.3551288 &  $-$0.054055 &  $-$   & 0.2589 & 0.0903 &80.6838.2884  &  0.3583390 &     0.000478 & 0.0707 & 0.2950 & 0.0599 \\
14.9463.846   &  0.2749961 &     0.070745 & 0.0470 & 0.1293 &  $-$   &80.6951.2395  &  0.3114197 &  $-$0.000583 & 0.0487 & 0.1821 & 0.0597 \\
14.9702.401   &  0.2754043 &  $-$0.184006 &  $-$   & 0.2349 & 0.0469 &80.6953.1751  &  0.3517135 &  $-$0.028083 & 0.0677 & 0.2699 & 0.0709 \\
15.10068.239  &  0.2976949 &  $-$0.033619 & 0.0325 & 0.1769 & 0.0398 &80.6957.409   &  0.4078831 &     0.000793 & 0.0537 & 0.1470 & 0.0452 \\
15.10072.918  &  0.2930774 &  $-$0.036838 &  $-$   & 0.2971 & 0.0649 &80.6958.1037  &  0.2752632 &  $-$0.034210 & 0.0355 & 0.2141 & 0.0392 \\
15.10311.782  &  0.3482903 &     0.001036 & 0.0440 & 0.2308 & 0.0450 &80.7072.1545  &  0.3583444 &  $-$0.001338 & 0.0268 & 0.1236 & 0.0263 \\
15.10313.606  &  0.2925053 &     0.064385 & 0.0806 & 0.1861 & 0.0597 &80.7072.2280  &  0.2786360 &  $-$0.047201 &  $-$   & 0.1486 & 0.0359 \\
15.11036.255  &  0.2869357 &     0.078243 & 0.1029 & 0.1030 &  $-$   &80.7319.1287  &  0.2978598 &     0.000852 & 0.0973 & 0.2210 & 0.0761 \\
15.11280.663  &  0.3299770 &  $-$0.000580 &  $-$   & 0.2853 & 0.0393 &80.7437.1678  &  0.2781401 &  $-$0.088902 &  $-$   & 0.1858 & 0.1289 \\
18.2361.870   &  0.3254790 &     0.029733 & 0.0661 & 0.2112 & 0.0650 &80.7440.1192  &  0.3430250 &  $-$0.001029 & 0.0582 & 0.1660 & 0.0663 \\
19.4188.1264  &  0.3702574 &     0.000911 & 0.0945 & 0.1690 & 0.0678 &80.7441.933   &  0.2733126 &     0.047918 & 0.0551 & 0.1684 & 0.0420 \\
19.4188.195   &  0.2828193 &     0.012127 & 0.0683 & 0.1752 & 0.0665 &81.8400.901   &  0.2739736 &     0.006469 & 0.0404 & 0.1651 &  $-$   \\
2.4787.770    &  0.3357061 &  $-$0.000461 & 0.0617 & 0.2749 & 0.0620 &81.8518.970   &  0.3085487 &  $-$0.014972 &  $-$   & 0.1192 & 0.0264 \\
2.5032.703    &  0.3077973 &  $-$0.006788 &  $-$   & 0.1864 & 0.0344 &81.8639.1749  &  0.3759094 &     0.044767 & 0.0459 & 0.2201 & 0.0363 \\
2.5148.1207   &  0.2946848 &  $-$0.035316 & 0.0611 & 0.1824 & 0.0664 &81.8759.832   &  0.3038356 &     0.074458 & 0.0377 & 0.2144 &  $-$   \\
2.5148.713    &  0.3212517 &  $-$0.000684 & 0.0318 & 0.1651 & 0.0333 &81.8879.1869  &  0.2975649 &  $-$0.000735 & 0.0614 & 0.1871 & 0.0660 \\
2.5266.3864   &  0.2793840 &     0.065535 & 0.0705 & 0.1768 &  $-$   &82.7920.1074  &  0.3386711 &     0.038043 & 0.0470 & 0.2152 &  $-$   \\
2.5271.255    &  0.4347510 &     0.001169 & 0.0658 & 0.1821 & 0.0541 &82.8049.746   &  0.2987354 &     0.071957 & 0.0675 & 0.2256 & 0.0482 \\
2.5876.741    &  0.2859854 &  $-$0.072205 &  $-$   & 0.2184 & 0.0347 &82.8286.1784  &  0.3285773 &  $-$0.000494 &  $-$   & 0.1893 & 0.0357 \\
3.6240.450    &  0.3836371 &  $-$0.001203 & 0.0641 & 0.2129 & 0.0583 &82.8407.312   &  0.3563439 &     0.000804 & 0.0371 & 0.2317 &  $-$   \\
3.6597.756    &  0.3086451 &  $-$0.000950 & 0.0304 & 0.1378 & 0.0367 &82.8408.1002  &  0.2962930 &     0.035408 & 0.0694 & 0.2299 & 0.0503 \\
3.6603.795    &  0.2765073 &  $-$0.017362 &  $-$   & 0.2527 & 0.0369 &82.8525.1980  &  0.2833982 &     0.081435 & 0.0355 & 0.2378 &  $-$   \\
3.6838.1298   &  0.3583389 &     0.000466 & 0.0646 & 0.3078 & 0.0563 &82.8526.1176  &  0.3238862 &     0.127004 & 0.0321 & 0.1293 & 0.0312 \\
3.6966.427    &  0.2972764 &     0.078251 & 0.0272 & 0.2148 &  $-$   &82.8765.1250  &  0.3049238 &  $-$0.048999 & 0.0422 & 0.2267 & 0.0538 \\
3.7088.623    &  0.3244180 &  $-$0.000642 & 0.0530 & 0.2485 & 0.0676 &82.8766.1305  &  0.2592516 &  $-$0.003018 &  $-$   & 0.0936 & 0.0518 \\
47.1521.589   &  0.3587236 &  $-$0.042303 & 0.0575 & 0.2303 & 0.0634 &9.4274.644    &  0.2636423 &  $-$0.122986 &  $-$   & 0.1529 & 0.0343 \\
5.4401.1018   &  0.2685325 &  $-$0.004754 & 0.0761 & 0.0830 & 0.0802 &9.4632.731    &  0.2775501 &  $-$0.006687 & 0.0310 & 0.1220 & 0.1233 \\
5.4528.1555   &  0.2752904 &  $-$0.010273 &  $-$   & 0.0908 & 0.0212 &9.4636.1979   &  0.3933065 &     0.000434 & 0.0254 & 0.2201 &  $-$   \\
5.4889.1060   &  0.2815683 &  $-$0.097085 &  $-$   & 0.1052 & 0.0309 &9.5117.617    &  0.3601338 &  $-$0.000491 & 0.0265 & 0.2328 & 0.0341 \\
5.5008.1902   &  0.3774283 &     0.073639 & 0.0436 & 0.1327 &  $-$   &9.5125.1018   &  0.2598093 &  $-$0.005181 & 0.0376 & 0.1075 & 0.0572 \\
5.5250.1501   &  0.3551757 &  $-$0.000513 & 0.0378 & 0.2143 & 0.0433 &9.5242.1032   &  0.2858950 &     0.037524 & 0.0879 & 0.1245 &  $-$   \\
5.5367.3432   &  0.2854189 &     0.029594 & 0.0370 & 0.1800 & 0.0367 &9.5479.852    &  0.3217303 &     0.000755 & 0.1144 & 0.1728 & 0.0827 \\
5.5368.1201   &  0.3365130 &  $-$0.005552 & 0.0287 & 0.2049 & 0.0554 &9.5481.746    &  0.3581725 &  $-$0.000506 & 0.0234 & 0.2063 & 0.0345 \\
5.5489.1397   &  0.2899676 &  $-$0.047824 &  $-$   & 0.1849 & 0.0719 &9.5606.348    &  0.2909644 &     0.071233 & 0.0226 & 0.1168 & 0.0209 \\
5.5497.3874   &  0.2680527 &  $-$0.018829 &  $-$   & 0.1566 & 0.0381 &              &            &              &        &        &        \\
\hline
\end{tabular}
\end{flushleft}
{\footnotesize
\underline {Notes:} 
The amplitude of the main pulsation component is denoted by 
$A_0$. The modulation amplitudes are $A_{+}$ and $A_{-}$, corresponding 
to the larger and smaller frequencies, respectively. Amplitudes are 
given in MACHO instrumental `b' magnitudes, frequencies in [d$^{-1}$]. 
The sign of the modulation frequency $f_{\rm BL}$ is positive, if 
$A_{-}<A_{+}$ and negative, if $A_{-}>A_{+}$. For BL2 stars $f_{\rm BL}$ 
stands for the average of the two modulation frequencies (see text 
for further details). This table appears also at CDS 
(http://cdsweb.u-strasbg.fr).
}
\end{table*}

The datasets used in the analysis are described in Sect.~2. The 
final statistics of classification are shown in Table~4. 
Additional details of the analysis on a star-by-star basis are 
given in Table~5. The result presented in Table~4 is based on 
the analysis of set \#2 comprising the full available two-color 
data. In comparison with the similar summary of A00 (their Table~7), 
we see that the most striking difference is the increase of the 
incidence rate of the BL stars by almost a factor of two. This 
change can be attributed to the following effects. 
\begin{itemize}
\item
{\it Longer time span:} +4 stars 
-- from the comparison of the detections in the `r' data. 
\item
{\it The better quality and higher amplitudes in the `b' data:} 
+28 stars -- from the comparison of the set \#2 `r' and `b' data.
\item
{\it Extending the allowed range of BL periods up to the length 
of the total time span:} +12 stars
-- based on the SAM statistics of the set \#2 results.
\item
{\it Employing SAM:} +9 stars
-- from the comparison of the set \#2 `b' and SAM analyses.
\item
{\it False detections in A00:} $-7$ stars 
-- due to the higher cutoff employed here for detection limit 
and because of the somewhat different classification scheme. 
\end{itemize}     

For the above reasons, we also find increases in the number 
of other types of variables. On the other hand, three stars, 
classified earlier by A00 as RR12 have been re-classified as 
MDM, or `mysterious double-mode'. Basic properties of these 
stars are summarized in Table~6. These objects have been 
re-classified mostly because of their somewhat extreme 
position in the color-magnitude diagram. Indeed, if we plot 
the derived average magnitudes and color indices on the 
color-magnitude diagram of Alcock et al. (2000a), the three 
MDM stars fall to the high-luminosity red edge of the region 
populated by RR1 stars. We recall that in the LMC,  
$\langle V \rangle = 19.35$ and $\langle V-R \rangle =0.22$, 
for the RR1 stars, with $V-R < 0.3$ for most of the variables 
(see Alcock et al. 2004). Although these objects are about 
1~mag fainter than the faintest first/second overtone double-mode 
Cepheids (see Soszy\'nski et al. 2000), it may still not be 
excluded that they are faint overtone Cepheids rather than 
bright and very red overtone RR Lyraes. It is also noted that 
there is a narrow overlap in the periods of these two classes 
of stars. One needs more accurate data to understand the status 
of these intriguing objects.

A slight change in the statistics of the above classification 
occurs if we consider double identifications due to field overlaps. 
We consider a star to be double-identified, if the following two 
conditions are satisfied simultaneously: (i) the simple distance 
derived from the co-ordinates is smaller than $2\times 10^{-3}$ 
degrees; (2) the difference between the periods is smaller than 
$10^{-5}$~d. We find altogether 52 double-identified objects. 
Among these there are 32 RR1-S, 8 RR01, 5 PC, 1 BL1 and 3 BL2. 
It is noted that except for 3 marginal cases of PC/S ambiguities, 
the classifications of the double-identified objects are consistent. 
The multiple identifications lead to the revised incidence rates 
of 3.5\% and 3.9\% for the BL1 and BL2 stars, respectively. 
We see that the changes are insignificant. 

%
%
\subsection{Properties of the BL and MC stars}
The basic data on the 99 BL stars are summarized in Table~7.
The description of the columns is given in the note added to 
the table. As mentioned, in the case of BL2 stars, the modulation 
frequency $f_{\rm BL}$ stands for $(f_{+}+f_{-})/2$, where 
$f_{+}$, $f_{-}$ denote the frequencies of the $A_{+}$ and 
$A_{-}$ modulation components and $f_1$ the main pulsation 
frequency. This type of averaging is justified, because from 
the $53$ BL2 variables there are only $7$ with  
$\delta f=|f_{+}+f_{-}-2f_1|>0.0001$, $4$ with  
$\delta f>0.0002$ and only one (star 15.10068.239) with 
$\delta f=0.00048$. Although this latter value might reflect 
statistically significant deviation from equidistant frequency 
spacing, the majority of the the variables are well within 
the acceptable deviations due to observational noise (see A00 
for numerical tests).

The BL variables have been selected on the basis of the classification 
scheme described in Sect.~4. In that scheme, MC-type variables 
(i.e., those with multiple close periods) have been excluded from the 
class of BL stars. One may wonder if these variables could also be 
considered as some subclass of the BL variables. We think that without 
knowing the underlying physics behind the BL phenomenon, it is up 
to one's preference if class MC is considered as subclass BL. Although 
here we opted not to take this choice, we note that there are five  
variables among the MC stars that show symmetric peaks of BL2-type 
within the multiple peaks. In Fig.~7 we show two representative cases 
when the presence of the BL2 structure is obvious. 
%
%
   \begin{figure}[h]
   \centering
   \includegraphics[width=70mm]{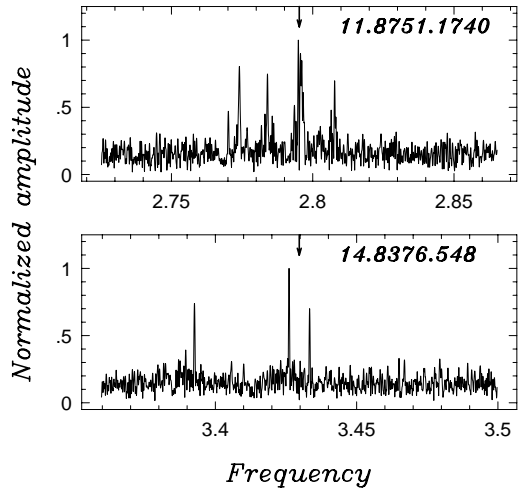}
      \caption{Examples of type MC variables with BL2 structures. 
      Arrow indicates the position of the (prewhitened) main 
      pulsation component. Spectra are computed by SAM. See text 
      for comments on the peak structures.} 
         \label{fig7}
   \end{figure}

In order to be more specific on all five stars containing structures of 
type BL2, below we give a short description of their frequency spectra.
In the following we use the notation $\nu_0$ for the main pulsation 
component.

\noindent
{\bf 11.8751.1740} --
The frequency spectrum shows some PC-type remnant at $\nu_0$, 
a symmetric pair of peaks around it and an additional component 
further away (see Fig.~7).

\noindent
{\bf 14.8376.548} --
The frequency spectrum shows a symmetric pair of closely spaced 
peaks around $\nu_0$ and an additional peak further away (see Fig.~7).

\noindent
{\bf 18.2357.757} --
Classified as BL2 from the SAM analysis of set \#1. From set \#2 we 
found an additional close component, therefore the final classification 
has become of MC.

\noindent
{\bf 19.4671.684} --
The frequency spectrum shows a dominantly BL2 structure, but there 
are additional peaks spred closely.

\noindent
{\bf 6.6697.1565} --
The frequency spectrum shows two BL2 structures superposed, one with 
large and another one with small frequency separations. The closer 
pair has also lower amplitudes.

 The remaining eight stars from the MC variables show distinct peaks 
without any apparent BL2-type peak structures. Representative examples 
of these variables are shown in Fig.~8. We see that these are indeed 
different from the ones generally classified as type BL.
%
%
%
   \begin{figure}[h]
   \centering
   \includegraphics[width=70mm]{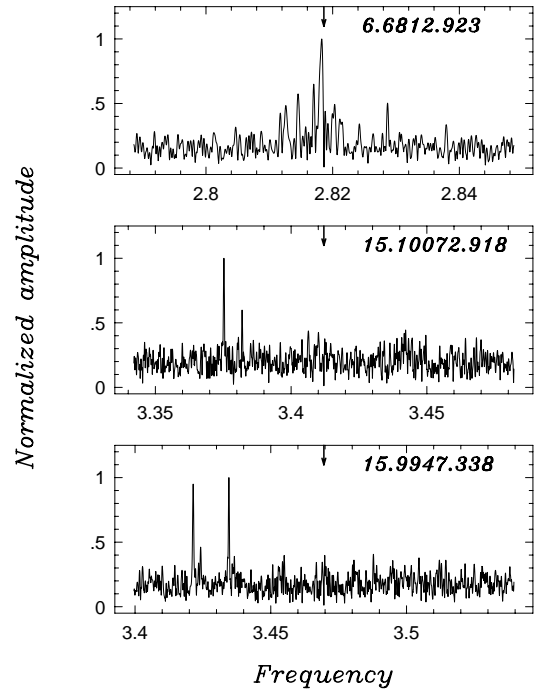}
      \caption{Examples of type MC variables without symmetric 
      peak structures of type BL2. Arrow indicates the position 
      of the (prewhitened) main pulsation component. 
      Spectra are computed by SAM.} 
         \label{fig8}
   \end{figure}

It is interesting to examine the distribution of the size and 
the position of the modulation components. For comparison, in 
Fig.~9 we show the distributions of the modulation amplitudes 
both for the RR1 and RR0 BL stars. It is seen that in both 
classes there are variables with very high modulation amplitudes, 
exceeding $0.1$~mag. This may correspond to $\sim 50$\% and  
$\sim 90$\% relative modulation levels for the RR0 and RR1 stars, 
respectively. In the other extreme, at low modulation amplitudes, 
we find cases near $0.01$--$0.02$~mag. From the size of the noise 
and the number of the data points, we expect about these 
amplitudes as the lowest ones to be detected in this dataset. 
It is important to note that current investigations by Jurcsik 
and co-workers (Jurcsik et al. 2005a) suggest that modulation 
may occur under $0.01$~mag among fundamental mode Galactic field 
Blazhko stars. It is clear that we need more accurate data on 
the LMC to check if these low modulation levels are more common 
also in other stellar systems.  

%
%
%
   \begin{figure}[h]
   \centering
   \includegraphics[width=70mm]{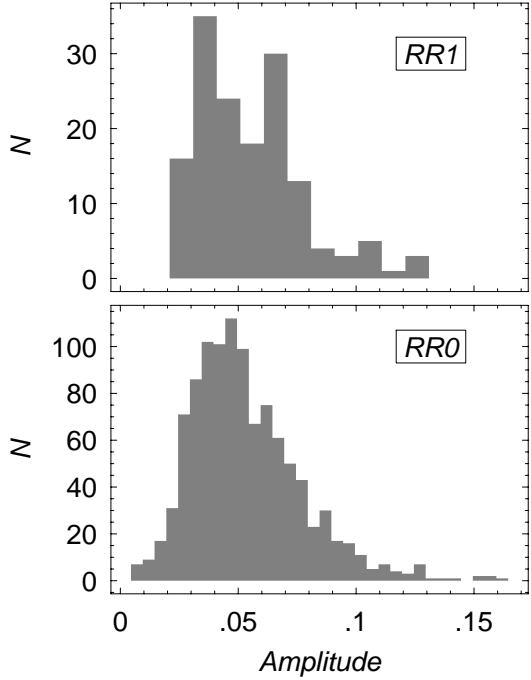}
      \caption{{\it Upper panel:} distribution of the `b' modulation 
      amplitudes of the 99 RR1 stars shown in Table~7. In the case 
      of BL2 stars the amplitudes of both side peaks are included.  
      {\it Lower panel:} as in the upper panel, but for the `V' 
      modulation amplitudes of the 731 RR0 stars of A03. 
      } 
         \label{fig9}
   \end{figure}

Concerning the relative positions of the modulation components, 
in Table~8 we show the number of cases obtained in the different 
datasets with the larger modulation component preceding the main 
pulsation component. We see that in the case of RR1 stars we have 
nearly 50\% probability that this happens. This result is basically 
independent of BL type, color and extent of the dataset. We recall 
that A03 derived 75\% for this ratio for the RR0 stars in the LMC. 
%
%
%
\begin{table}
\caption{Positions of the larger modulation amplitudes}
\begin{flushleft}
\begin{tabular}{clccc}
\hline\hline
{\rm Type} & {\rm Color} & $N_+$ & $N_{\rm tot}$ & $N_+/N_{\rm tot}$ \\
\hline
BL1   & {\rm `r'}    & 12   & 32 & 0.38 \\
      & {\rm `b'}    & 18   & 44 & 0.41 \\
      & {\rm `SAM'}  & 21   & 46 & 0.46 \\
BL2   & {\rm `r'}    & 13   & 22 & 0.59 \\
      & {\rm `b'}    & 21   & 46 & 0.46 \\
      & {\rm `SAM'}  & 24   & 53 & 0.45 \\
\hline
\end{tabular}
\end{flushleft}
{\footnotesize
\underline {Notes:} 
Dataset \#2 is used; $N_+=$ number of cases when the larger 
modulation amplitude has a frequency greater than that of 
the main pulsation component; $N_{\rm tot}=$ total number 
of BL stars identified in the given color.
}
\end{table}    

The present study has largely extended the range of the BL periods 
known for RR1 stars. The distributions of the modulation frequencies 
for the RR1 and RR0 stars are shown in Fig.~10. Although the 
sample is much smaller for the RR1 stars, it is clear that there 
are more stars among them with short modulation periods than among 
the RR0 stars. Nevertheless, we should mention that Jurcsik et al. 
(2005a, b) found very short modulation periods also among RR0 stars, 
albeit in the Galactic field (i.e., SS~Cnc and RR~Gem with 
$P_{\rm BL}=5.3$ and $7.2$~d, respectively). It is also noted 
that the considerable surplus in the RR1-BL stars with long modulation 
periods can be attributed only partially to our lower limit set on 
$f_{\rm BL}$. We find that among the $34$ stars contributing to the 
first bin in Fig.~10 for the plot of RR1 stars, there are only $11$ 
with $1.0/T<f_{\rm BL}<1.5/T$ and $16$ with $1.0/T<f_{\rm BL}<2.0/T$. 
This implies a significant peak in the distribution function for 
$f_{\rm BL}<0.005$ even with the omission of these objects with 
very long modulation periods. The lack of such a peak for the RR0 
stars might indicate a difference in this respect between the RR0 
and RR1 BL stars, but, due to the low sample size for the RR1 stars, 
we would caution against jumping to this conclusion.     
%
%
%
   \begin{figure}[h]
   \centering
   \includegraphics[width=70mm]{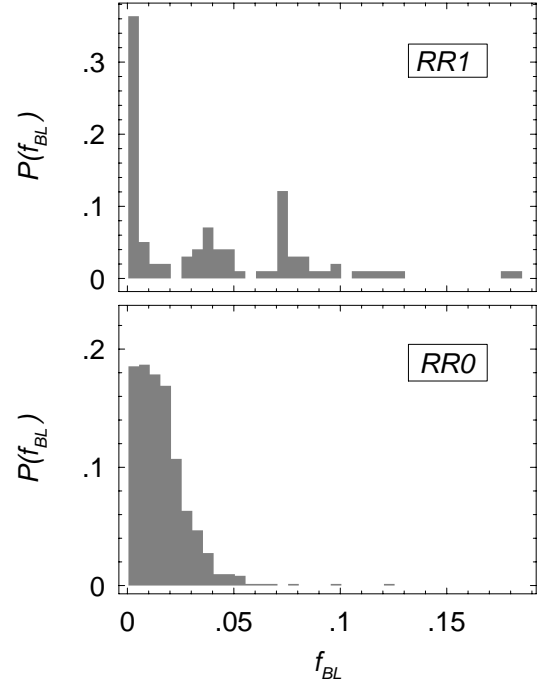}
      \caption{{\it Upper panel:} distribution of the modulation 
      frequencies of the 99 RR1 stars shown in Table~7.   
      {\it Lower panel:} as in the upper panel, but for the  
      modulation frequencies of the 731 RR0 stars of A03. 
      } 
         \label{fig10}
   \end{figure}
%

%
%
\subsection{Miscellaneous variables}
In Table~9 we list the properties of the variables showing frequency 
spectra difficult to classify. We see that in most cases the detections 
are barely above the noise level. Therefore, we cannot exclude that at 
least some of these variables will turn out to be single-periodic RR1 
stars, when more accurate data will be available. In the column 
``Comments'' we show the period ratios in cases when some sort of 
double-mode pulsation can be suspected (without suggesting that those 
variables are indeed of double-mode ones).

%
%
\begin{table*}
\caption{List of miscellaneous stars}
\begin{flushleft}
\begin{tabular}{lccrcl}
\hline\hline
MACHO ID&$\nu_{0}$&$\nu_{MI}$& SNR &Source&Comments \\
\hline
10.3916.849  & 3.1389921 & 1.2070180 & 6.61 & SAM &                           \\  
11.8745.899  & 3.3176002 & 4.7867033 & 6.79 & SAM & $\nu_{0}/\nu_{MI}=0.6931$ \\
13.5719.713  & 3.0220610 & 3.9155071 & 6.96 & SAM & $\nu_{0}/\nu_{MI}=0.7718$ \\
18.2717.787  & 2.8643249 & 4.9297051 & 6.55 & SAM & $\nu_{0}/\nu_{MI}=0.5810$ \\
18.3202.956  & 2.2855398 & 5.1908177 & 6.96 & r   &                           \\
19.3575.541  & 3.3698536 & 0.5856270 & 7.91 & SAM &                           \\
3.7451.484   & 3.3048194 & 2.7630866 & 7.85 & b   & $\nu_{MI}/\nu_{0}=0.8361$ \\
80.6348.2470 & 2.9571188 & 1.2069597 & 6.65 & SAM &                           \\
80.7558.650  & 2.7828759 & 0.8240633 & 6.99 & r   &                           \\
81.8519.1395 & 2.9463890 & 0.8372214 & 6.96 & r   &                           \\
9.5004.750   & 3.2877325 & 0.1604569 & 7.42 & SAM &                           \\
9.5122.363   & 3.1033541 & 4.1679579 & 6.82 & SAM & $\nu_{0}/\nu_{MI}=0.7446$ \\
9.5241.382   & 3.1777911 & 1.3826455 & 6.79 & SAM &                           \\
\hline
\end{tabular}
\end{flushleft}
{\footnotesize
\underline {Notes:} 
$\nu_{0}$ denotes the frequency (in [d$^{-1}$]) of the main pulsation 
component, $\nu_{MI}$ stands for the frequency of the miscellaneous 
component with SNR given in the next column. Interesting period 
ratios are given in the comment lines. 
}
\end{table*}
%
%

%
%
\section{Conclusions}
This work has been motivated by an earlier result of Alcock et al. 
(2000) on the low incidence rate of the first overtone Blazhko 
RR~Lyrae (RR1-BL) stars in the LMC. They derived a rate of 
4\% for these stars, which is a factor of three lower than that 
of the fundamental mode (RR0) stars (see Alcock et al. 2003). 
Soszy\'nski et al. (2003) obtained a somewhat higher rate 
of 6\% from the OGLE database. However, Alcock et al. (2000) used 
MACHO `r' data that have somewhat lower signal-to-noise ratio for 
the objects of interest, and, in addition, the data analyzed then, 
spanned a shorter time base than the ones available now. This 
suggests that the incidence rate could be higher than the one 
deciphered earlier. In a full utilization of all available data 
we employed a spectrum averaging method that enabled us to 
increase the detection rate by 18\% in comparison to the best 
single color (i.e., `b') rates. 

The main conclusion of this paper is that indeed, the incidence 
rate of the RR1-BL stars in the LMC is higher than previously 
derived from nearly the same database. Even though the rate is 
7.5\% now, it is still significantly lower than that of the RR0 
stars. This latter rate is 12\%, that we also expect to increase 
slightly when the same method as the one used in this paper on 
RR1 stars is to be employed also on RR0 stars. This means, that, 
at least for the LMC, we can treat as a well-established fact 
that RR1-BL stars are significantly less frequent than their 
counterparts among the RR0 stars. It is also remarkable that 
after filtering out the main pulsation component, the highest 
amplitude peak appears with equal probability at both sides of 
the frequency of the main pulsation component. We recall that 
in the case of RR0 stars there is a 75\% preference toward the 
higher frequency side. The smallest modulation amplitudes 
detected for RR1 stars are near $0.02$~mag. For the larger 
sample of RR0 stars this limit goes down to $0.01$~mag. The 
relative size of the modulation may reach $90$\% for RR1 stars, 
whereas the same limit is only $50$\% for the RR0 stars. 
Furthermore, RR1-BL stars have a relatively large population 
of short-periodic ($P_{\rm BL}<20$~d) variables, which region 
is nearly empty for the RR0 stars in the LMC. 

Since the underlying physical mechanism of the BL phenomenon 
is still unknown, it is also unknown if these (and other) 
observational facts will put strong (or even any) constraints 
on future theories. Nevertheless, since these statistics are 
based on large samples, their significance is high and surely 
cannot be ignored in the theoretical discussions. For example, 
we note that the very existence of BL1 stars puts in jeopardy 
the present forms of both currently available models (magnetic 
oblique rotator/pulsator by Shibahashi 2000; non-radial resonant 
model by Nowakowski \& Dziembowski 2001; see however Dziembowski 
\& Mizerski 2004).

\begin{acknowledgements}
This paper utilizes public domain data obtained by the MACHO Project, 
jointly funded by the US Department of Energy through the University 
of California, Lawrence Livermore National Laboratory under contract 
No. W-7405-Eng-48, by the National Science Foundation through the 
Center for Particle Astrophysics of the University of California 
under cooperative agreement AST-8809616, and by the Mount Stromlo 
and Siding Spring Observatory, part of the Australian National 
University. The support of grant T$-$038437 of the Hungarian 
Scientific Research Fund (OTKA) is acknowledged.
\end{acknowledgements}

\end{document}